\documentclass{article}
\usepackage{spconf,amsmath,graphicx}
\usepackage{amssymb}
\usepackage{xcolor}
\usepackage{booktabs}
\usepackage[hidelinks]{hyperref}
\usepackage{enumitem}
\setlist{nosep, leftmargin=14pt}

\usepackage{mwe} 

\title{Self-supervised OCT Image Denoising with Slice-to-Slice Registration and Reconstruction}
\name{\begin{tabular}{@{}c@{}}
Shijie Li$^{\dagger}$\thanks{Corresponding author: shijie.li@nyu.edu.} \qquad 
Palaiologos Alexopoulos$^{\star}$ \qquad 
Anse Vellappally$^{\star}$ \qquad
Ronald Zambrano$^{\star}$ \\
Wollstein Gadi$^{\star}$ \qquad 
Guido Gerig$^{\dagger}$
\end{tabular}}
\address{$^{\dagger}$ Computer Science and Engineering, New York University, Brooklyn, NY, USA. \\$^{\star}$Department of Ophthalmology, New York University, New York, NY, USA}
\begin{document}
%
\maketitle
\begin{abstract}
Strong speckle noise is inherent to optical coherence tomography (OCT) imaging and represents a significant obstacle for accurate quantitative analysis of retinal structures which is key for advances in clinical diagnosis and monitoring of disease. Learning-based self-supervised methods for structure-preserving noise reduction have demonstrated superior performance over traditional methods but face unique challenges in OCT imaging. The high correlation of voxels generated by coherent A-scan beams undermines the efficacy of self-supervised learning methods as it violates the assumption of independent pixel noise. We conduct experiments demonstrating limitations of existing models due to this independence assumption. We then introduce a new end-to-end self-supervised learning framework specifically tailored for OCT image denoising, integrating slice-by-slice training and registration modules into one network. An extensive ablation study is conducted for the proposed approach. Comparison to previously published self-supervised denoising models demonstrates improved performance of the proposed framework, potentially serving as a preprocessing step towards superior segmentation performance and quantitative analysis. \href{https://github.com/CJLee94/Slice2Slice.git}{Code is publicly available}.
\end{abstract}

\section{Introduction}
\label{sec:intro}

Optical Coherence Tomography (OCT) imaging is the most important imaging modality for retinal studies,  yet the challenge of speckle noise significantly impedes precise quantitative assessment of retinal structures and thus may pose limitations to diagnosis, monitoring of pathology and treatment decisions. Conventional structure-preserving denoising such as median filtering or BM3D/BM4D \cite{dabov2007bm3d, meggioni2013bm4d} provide insufficient noise reduction, tend to blur images and are prone to artefacts. Learning-based methods using Convolutional Neural Networks (CNNs) depend heavily on sets of clean images for training, a difficult  requirement in medical imaging due to the scarcity of such reference scans.

Recent developments, such as the Noise2Noise model by Lehtinen et al. \cite{lehtinen2018noise2noise}, have shown promise by training on multiple images with identical content but independent noise. This approach, applied to repeated OCT scans \cite{gisbert2020self}, effectively preserves fine details while removing noise. Nonetheless, obtaining repeated scans in medical settings is often impractical, and motion and minor deformations between scans, especially in retinal OCT images, necessitate intense preprocessing for co-registration. Furthermore, self-supervised denoising methods \cite{krull2019noise2void, batson2019noise2self, huang2021neighbor2neighbor}, proposed as alternatives, show limited success in OCT denoising. These methods typically overlook the strong inter-voxel correlations in OCT scans, which violate basic assumptions of these methods which leads to suboptimal noise modeling and reduced denoising effectiveness. Techniques like \cite{oguz2020self}, which average co-registered neighboring slices, and \cite{rico2022real}, which train networks using noisy slices as input and averaged neighboring slices as targets, attempt to mitgate this issue. But our experiment indicate that while averaging partly reduces noise correlation between noise of input and output, it does not entirely eliminate it.
\begin{figure}[t]
    \centering
    \includegraphics[width=0.5\textwidth]{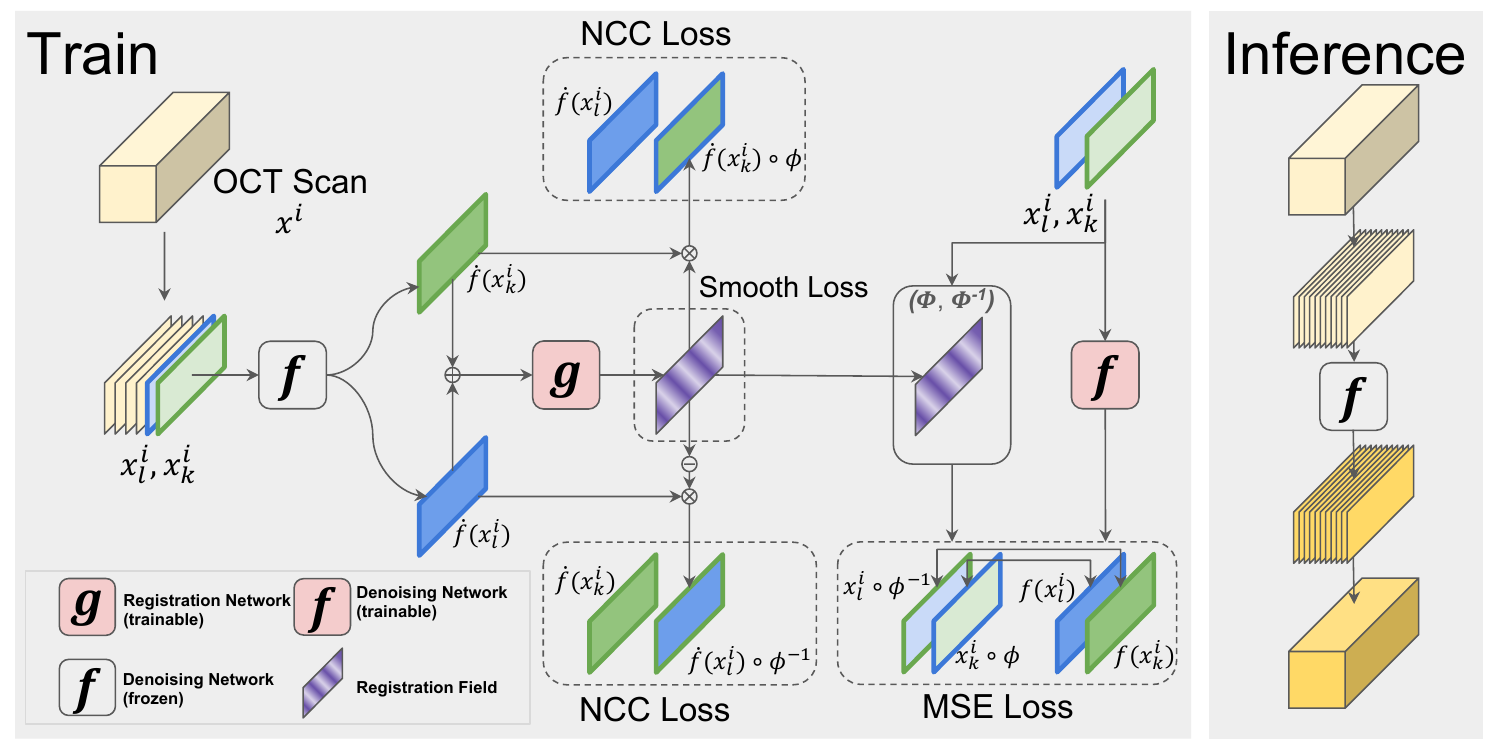}
    \caption{Flowchart illustrating the training and inference processes within the self-supervised framework for OCT image denoising.}
    \label{fig:pipeline}
\end{figure}
In response to these challenges, our study focuses on elucidating the so far underexplored inherent speckle noise correlations among OCT voxels and their implications on denoising performance. We introduce an new end-to-end self-supervised learning framework specifically designed for OCT image denoising. In addition, reducing the need for extensive preprocessing steps on training data, our method enhances the efficiency and applicability of OCT denoising across larger datasets. By acknowledging the unique properties of OCT scans, we demonstrate that our framework achieves superior denoising results while preserving subtle details. 

\section{Theoretical Background}
In single-image supervised denoising, a regression model \( f_{\theta} \) (such as a Convolutional Neural Network) is trained using pairs of corrupted input and clean target images \((\mathbf{x}_i, \mathbf{y}_i)\). The training process aims to solve the optimization problem:

\begin{equation}
    \min_\theta \sum_i \mathcal{L}(f_\theta(\mathbf{x}_i), \mathbf{y}_i).
\end{equation}

Here, \(\mathcal{L}(.)\) is a function that quantifies the similarity between the predicted and target outputs. The Noise2Noise framework, introduced by Lehtinen et al. \cite{lehtinen2018noise2noise}, proposes a significant modification. Theoretically, in the presence of an infinitely large dataset, the optimal parameters for the model can also be achieved by minimizing the loss between the model's predictions and a second set of corrupted images \(\mathbf{\hat{y}}_i\), rather than clean targets:

\begin{equation}
    \min_\theta \sum_i \mathcal{L}(f_\theta(\mathbf{x}_i), \mathbf{\hat{y}}_i).
\end{equation}

This approach depends on two critical conditions: the expectation \(E(\mathbf{\hat{y}}_i)\) must equal the clean image \(\mathbf{y}_i\), and the noise in \(\mathbf{\hat{y}}_i\) must be independent of the clean signal in \(\mathbf{y}_i\). 
\section{methodology}
\subsection{Motivation}
\begin{figure}[t]
    \centering
    \includegraphics[width=0.5\textwidth]{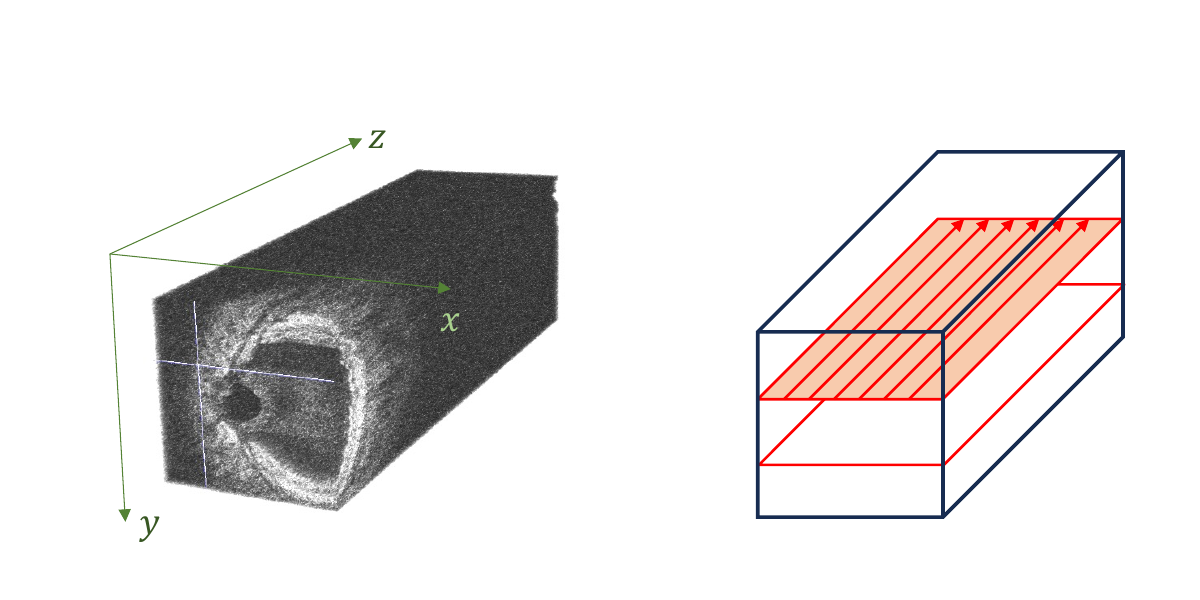}
    \caption{Left: A 3D-rendered OCT scan at $200\times1024\times200$ resolution. Right: OCT acquisition method, with coherent light scanned over the xy-plane and high noise correlation along the z-axis A-scans. Adjacent xy-plane en face or C-scan slices, with their correlated noise, are unsuitable as training pairs for denoising.}
    \label{fig:corr_oct}
\end{figure}

Our framework is motivated by the similarity observed in neighboring B-scan slices of 3D OCT images, making the Noise2Noise approach applicable as it treats these slices as noisy representations of the same scene. However, OCT's speckle noise correlation poses a challenge, as incorrect input-target pairing could result in the model learning noise characteristics instead of the clean signal(shown in Figure \ref{fig:corr_oct}). This necessitates OCT-specific denoising methods.

Huang et al.\cite{huang2021neighbor2neighbor} highlight that discrepancies in clean signals can impact Noise2Noise's effectiveness. To optimize performance, image alignment prior to training is critical, thus our framework integrates a trainable image registration network following Balakrishnan et al. \cite{balakrishnan2019voxelmorph}.



Additionally, we recognize a challenge with the method of \cite{balakrishnan2019voxelmorph}: its performance is limited to process noisy images with low signal-to-noise ratios. This observation further motivates our framework's design to add a pre-denoising step before the image registration network.
\subsection{Slice2Slice Denoising Network}
Our framework targets OCT denoising challenges by using neighboring B-scans to minimize noise correlation, and parallel training of VoxelMorph for scan alignment. We also include a preliminary denoising step to improve VoxelMorph's registration accuracy, which helps address image misalignment caused by noise without impacting the primary denoising process. 
\subsubsection{Image registration network}
\begin{figure*}[t]
    \centering
    \includegraphics[width=\textwidth]{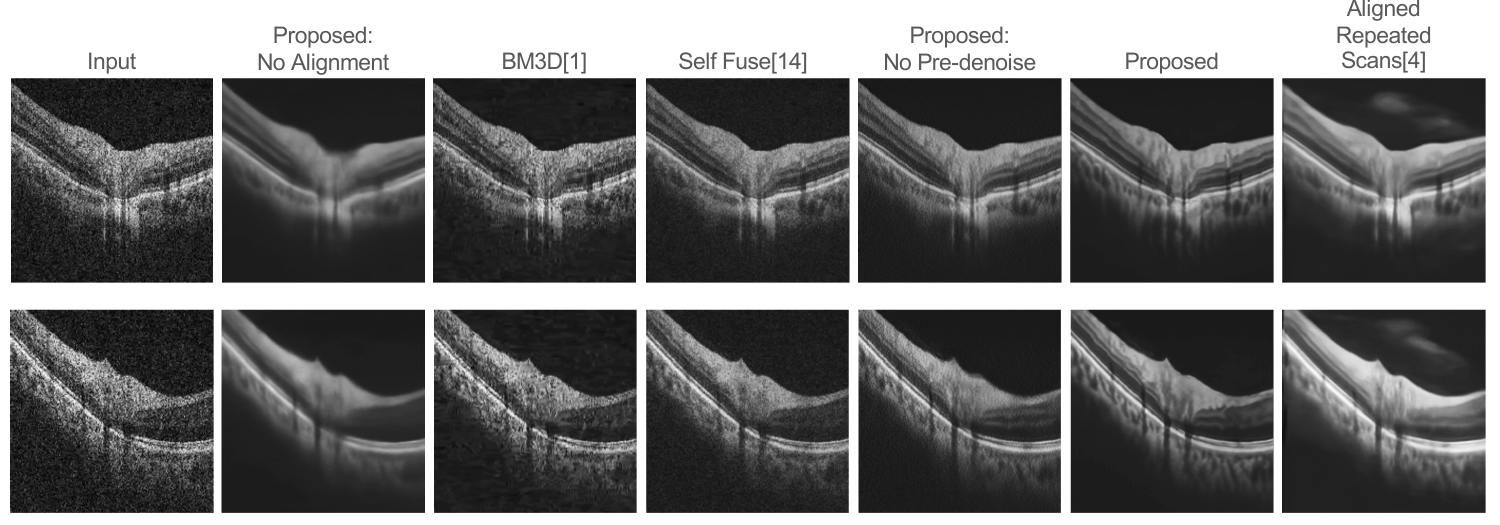}
    \caption{Sequentially from left to right: original noisy B-scan images; results from training with our pipeline excluding the image registration module; denoising outcomes using BM3D \cite{dabov2007bm3d}; results from Self Fuse \cite{rico2022real}; images denoised by our model trained without pre-denoising before image registration; images denoised by our complete proposed pipeline; and images denoised by the method described in \cite{gisbert2020self}.}
    \label{fig:results}
\end{figure*}

In Optical Coherence Tomography (OCT) imaging, while neighboring B-scan slices generally show high similarity, discrepancies often arise due to factors like eye movements or significant changes in retinal structure. These misalignments can result in blurring in denoising models as evidenced in Section \ref{sec:exp_alignment} and Figure \ref{fig:results}. To address this, our approach processes adjacent pairs of B-scans (denoted as $\mathbf{x}^i_k, \mathbf{x}^i_l$) from the same OCT scan ($\mathbf{x^i}$), concatenating them along the channel axis. These pairs are then input into a UNet-based structure \cite{ronneberger2015u} to produce a displacement field ($\mathbf{u}$). Then, the registration field $\phi=\mathbf{I}+\mathbf{u}$ is formed, where $\mathbf{I}$ is the identity transform. This field effectively aligns one B-scan slice with another. Our spatial transformation module utilizes $\phi$ to align $\mathbf{x}^i_l$ with $\mathbf{x}^i_k$, resulting in an aligned output $\mathbf{\hat{x}}^i_l$. We also employ a bidirectional approach to generate a correspondingly aligned $\mathbf{\hat{x}}^i_k$. This bidirectional alignment, a novel aspect of our network, allows us to effectively train the model with limited data, in line with the methods proposed in \cite{calvarons2021improved}. As shown later, such integration of the Voxelmorph module not only enhances alignment accuracy but also improves the overall performance of the denoising process in our network. Our loss function for the image registration network becomes
\begin{equation}
    \mathcal{L}_{\text{NCC}}(\mathbf{x}^i_k, \mathbf{\hat{x}}^i_l) + \mathcal{L}_{\text{NCC}}(\mathbf{x}^i_l, \mathbf{\hat{x}}^i_k) + \lambda\mathcal{L}_{\text{smooth}}(\mathbf{u}), 
\end{equation}
and integrates two components: the Normalized Cross Correlation (NCC) for consistency between the two images, and a smoothness regularization which is the $L_2$ norm of the gradient of the displacement field $\mathbf{u}$. Here, $\lambda$ is a hyperparameter that controls the weight of the smoothness regularizer in the loss function. The latter is crucial for ensuring the deformation field's smoothness, an aspect particularly emphasized in Section \ref{sec:exp_alignment}. Given the minor deformations and noisy nature of OCT images, a stronger emphasis on regularization is warranted compared to the method's original application in anatomical imaging.

\subsubsection{Self supervised denoising network}
In our approach, we opt not to use the aligned B-scan slices directly as targets for the denoising network. A direct use would involve images subject to bilinear interpolation which would alter its noise characteristics. Instead, we leverage the displacement field from the image registration network and apply nearest neighbor interpolation to preserve the original noise distribution. During training, we use four types of images: the original B-scans ($\mathbf{x}^i_k$, $\mathbf{x}^i_l$) and their aligned counterparts ($\mathbf{\Bar{x}}^i_k$, $\mathbf{\Bar{x}}^i_l$) obtained using nearest neighbor interpolation. To avoid confusion with previous notations, $\mathbf{\Bar{x}}^i_k$ and $\mathbf{\Bar{x}}^i_l$ specifically denote the aligned noisy images.

For ease of explanation, let us represent the denoising network with the function $f(.)$. The network's training involves minimizing a loss function, defined as $L=\|f(\mathbf{x}^i_k)-\mathbf{\Bar{x}}^i_l\|^2_2+\|f(\mathbf{x}^i_l)-\mathbf{\Bar{x}}^i_k\|^2_2$. 


\begin{figure}[t]
    \centering
    \includegraphics[width=0.5\textwidth]{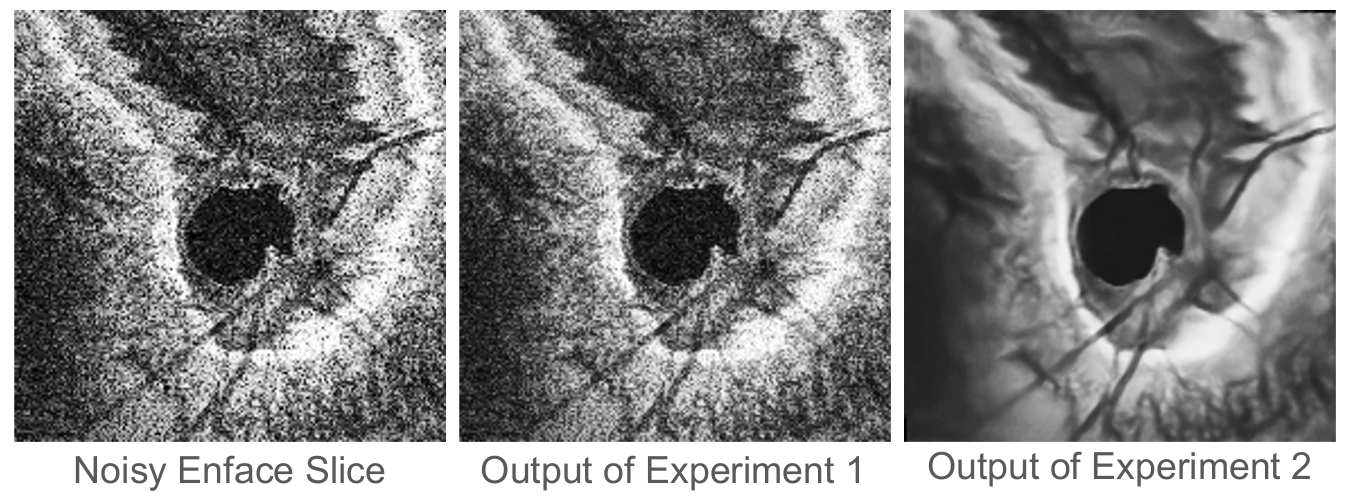}
    \caption{Experimental demonstration of how noise correlation between input and target image pairs impacts the efficacy of learning-based denoising models trained with the Noise2Noise\cite{lehtinen2018noise2noise} framework.}
    \label{fig:noise_corr}
\end{figure}
\section{Experiments}
In this section, we describe the setup of our training and testing datasets and outline a series of experiments designed to evaluate the proposed denoising model. These include a test of the impact of noise correlation on model performance, an ablation study examining the individual components of our image registration module, and comparisons with other OCT denoising methods such as \cite{dabov2007bm3d, meggioni2013bm4d, huang2021neighbor2neighbor, batson2019noise2self, rico2022real}. We perform these comparisons under scenarios involving training with both single and multiple OCT scans.
\subsection{OCT Data}
\label{subsec:OCT}
As discussed previously, our training does not rely on repeated scans of the same imaged structure but only requires a single 3D OCT scan or multiple  scans of different subjects but with similar noise characteristics. We employ Cirrus HD-OCT ONH scans at a voxel resolution of \(200\times1024\times200\) and dimensions of \(6\times2\times6\) mm³. The training set contains 20 scans, and the testing set has 9 scans.

To address the scarcity of clean OCT images and evaluate denoising algorithms against OCT's noise correlation, we employed simulated datasets, specifically designed to mimic OCT's noise patterns, using the IXI brain scan MRI dataset\footnote{https://brain-development.org/ixi-dataset/} as a baseline. Our experiments use 120 MRI images for training and 60 for testing.

\subsection{Evaluation Metrics}
We use the peak signal to noise ratio (PSNR) and structural similarity index measure (SSIM) calculated with the predicted denoised image and the groundtruth in the simulated dataset for quantitative evaluation of the model performance.
\subsection{Effect of Noise Correlation on Denoising\label{sec:verification}}
We conducted two experiments to determine the impact of noise correlation on the performance of the denoising model.

\noindent\textit{Experiment 1}: Trained with neighboring "en face" x-y slices ($200\times 200$ voxels) from the same OCT scans, the model failed to effectively denoise, as shown in the middle of Figure \ref{fig:noise_corr}, due to strong voxel correlations. As discussed previously in \ref{subsec:OCT}, subsequent x-directions of enface x-y slices are subject to pixel-by-pixel correlation, thus violating the basic assumption of our denoising method.

\noindent\textit{Experiment 2}: Using neighboring x-y slices from two co-registered OCT scans results in successful noise reduction, as indicated to the right of Figure \ref{fig:noise_corr}, demonstrating the importance of independent voxel sets for training.



\subsection{Ablation Study of Slice Alignment\label{sec:exp_alignment}}
Although neighboring OCT scan slices are generally similar, significant structural differences can occur, especially near the optical nerve head area or due to eye movements during scanning (see also horizontal line shifts in Fig.~\ref{fig:noise_corr}). To address this, we compared denoising models trained on both aligned and unaligned B-scans. Alignment was performed using ANTs~\cite{avants2009advanced} for co-registering slices. Our results (Figure \ref{fig:results}) show that unaligned B-scans produce blurriness in areas with structurally variability.

\subsection{Ablation Study of Pre-Denoising\label{sec:exp_predn}}
In the previous subsection, we identified that misalignment between input and target B-scans impairs the denoising model's capacity to accurately restore details such as blood vessels. This finding suggests that image noise might similarly affect the performance of the image registration module. To explore this, we hypothesize that applying denoising prior to image registration could improve the overall effectiveness of the denoising model. Testing was thus conducted to evaluate whether pre-denoising enhances image registration accuracy, thereby contributing to more effective model training.

\subsection{Comparison with other Methods}
We compare our model with other existing self-supervised denoising models including \cite{dabov2007bm3d, meggioni2013bm4d, batson2019noise2self, huang2021neighbor2neighbor, rico2022real}. We evaluate and compare them qualitatively and quantitatively by calculating the PSNR and SSIM metrics. The results are shown in the table.
\begin{table}[ht]
    \centering
    \begin{tabular}{lcc}
    \toprule
       \textbf{Method}  & \textbf{PSNR}$\uparrow$ & \textbf{SSIM}$\uparrow$ \\
       \midrule
        BM3D \cite{dabov2007bm3d} & 23.3 & 0.272 \\
        BM4D \cite{meggioni2013bm4d} & 24 & 0.298 \\
        Neighbor2Neighbor \cite{huang2021neighbor2neighbor} & 15.7 & 0.0785 \\
        Noise2Self \cite{batson2019noise2self} & 14.8 & 0.0301\\
        Self Fuse \cite{rico2022real}& 21.4 & 0.231 \\
        \midrule
        Proposed (No Pre-denoise) & 22.9 & 0.264 \\
        Proposed & $\mathbf{25.0}$ & $\mathbf{0.390}$ \\
    \bottomrule
    \end{tabular}
    \label{tab:my_label}
\end{table}


\section{Conclusion and Discussion}

Whereas developers often strive to develop image processing methodologies which are generic to the type of input data, imaging modalities such as OCT demonstrate that detailed knowledge on acquisition technology is necessary to achieve expected results. We present learning-based self-supervised denoising for 3D OCT imaging which, unlike previously published models, does not require training sets of repeated scans and can even be trained on single 3D OCT images. The proposed integration of training for noise reduction plus slice alignment to compensate for eye movements into a single workflow is seen to be a novel contribution, with the efficacy of each model and component also tested via ablation. Qualitative assessment of results and calculation of PSNR and SSIM metrics demonstrate a strong level of noise reduction while preserving detailed structures such as blood vessels and the pattern of retinal layers, both key elements for OCT-based diagnosis and monitoring of retinal pathology.

We see the lack of publicly available benchmark datasets based on clinically relevant ground truth labeling as a limitation of the current comparisons as shown here.  Such benchmarks, for example retinal layer measurements across the whole 3D scan or pore to beam structural analysis of the lamina cribrosa, may much better elucidate if advanced image processing may lead to progress in research and potentially improved clinical workflows, thus benefitting patients.

\newpage 

\section{Compliance with ethical standards}
\label{sec:ethics}

\textbf{Human data statement:}
The institutional review board and ethics committee at New York University (NYU) approved the study methods and data collection. The study followed the tenets of the Declaration of Helsinki and was conducted in compliance with the Health Insurance Portability and Accountability Act. Informed consent was obtained from all patients.

\section{Acknowledgements}
This work is supported by the grants NIH NIBIB R01EB021391, NIH 1R01EY030770-01A1, NIH-NEI 2R01EY013178-15,  and the New York Center for Advanced Technology in Telecommunications (CATT).

\bibliographystyle{IEEEbib}
\bibliography{strings,refs}

\end{document}